\documentclass{aa}

\usepackage{graphicx}
\usepackage{txfonts}
\usepackage{color,xspace}
\usepackage{enumitem}

\newcommand{\degrees}{\ensuremath{^\circ}}

\begin{document}

\titlerunning{Ubiquitous interaction signs in prolate rotators}
   \title{Ubiquitous signs of interactions in early-type galaxies with prolate rotation }

   \author{Ivana Ebrov{\'a} \inst{1}\thanks{\email{ebrova.ivana@gmail.com}}
          \and
          Michal B{\'i}lek \inst{1,2}
          \and 
          Ana Vudragovi{\'c}\inst{3}
          \and
          Mustafa~K. Y{\i}ld{\i}z\inst{4,5}
          \and
          Pierre-Alain Duc\inst{2}
          }

  \institute{Nicolaus Copernicus Astronomical Center, Polish Academy of Sciences, Bartycka 18, 00-716 Warsaw, Poland
        \and
             Universit{\'e} de Strasbourg, CNRS, Observatoire astronomique de Strasbourg (ObAS), UMR 7550, 67000 Strasbourg, France
        \and 
             Astronomical Observatory, Volgina 7, 11060 Belgrade, Serbia
        \and
             Astronomy and Space Sciences Department, Science Faculty, Erciyes University, Kayseri, 38039 Turkey
        \and
             Erciyes University, Astronomy and Space Sciences Observatory Applied and Research Center (UZAYB\.{I}MER), 38039, Kayseri, Turkey
             }

   \date{Received 17 February 2021; accepted 9 March 2021}


  \abstract
   {
A small fraction of early-type galaxies (ETGs) show prolate rotation, i.e. they rotate around their long photometric axis. 
In simulations, certain configurations of galaxy mergers are known to produce this type of rotation. 
}
   {
We investigate the association of prolate rotation and signs of galaxy interactions among the observed galaxies.
}
   {
We collected a sample of 19 nearby ETGs with distinct prolate rotation from the literature  and inspected their ground-based deep optical images for interaction signs\,--\,18 in archival images and one in a new image obtained with the Milankovi\'c telescope.
}
   {
Tidal tails, shells, asymmetric/disturbed stellar halos, or on-going interactions are present in all the 19 prolate rotators.
Comparing this with the frequency of tidal disturbance among the general sample of ETGs of a roughly similar mass range and surface-brightness limit, we estimate that the chance probability of such an observation is only 0.00087. 
We also found a significant overabundance of prolate rotators that are hosting multiple stellar shells.
The visible tidal features imply a relatively recent galaxy interaction.
That agrees with the Illustris large-scale cosmological hydrodynamical simulation, where prolate rotators are predominantly formed in major mergers during the last 6\,Gyr. 
In the appendix, we present the properties of an additional galaxy, NGC\,7052, a prolate rotator for which no deep images are available, but for which an HST image revealed the presence of a prominent shell, which had not been reported before.
}
   {}

               
   \maketitle
%

\section{Introduction}

Spiral galaxies, as well as most of the early-type galaxies (ETGs), rotate around their minor axis. 
A fraction of ETGs shows a significant misalignment between the photometric and kinematic axes. 
We say that a galaxy has prolate rotation, when the misalignment is close to 90\degrees, i.e. the galaxy appears to be rotating more around the major morphological axis.
This type of rotation is also often called “minor-axis rotation” since the gradient of the mean line-of-sight velocity is detected along the projected optical minor axis.
 
Minor-axis rotation was originally detected in one-dimensional (long-)slit spectroscopic data. 
Integral field spectroscopy (IFS), i.e. two-dimensional observations, reveals more convincing cases of prolate rotation, and IFS surveys enable estimates of the frequency of this phenomenon. 
In the ATLAS$^{3{\rm D}}$ project \citep{a3d1}, a complete volume and magnitude-limited sample of 260 ETGs, 3\,\% of galaxies show prolate rotation \citep{a3d2}. 
In the diameter-selected Calar Alto Legacy Integral Field Area \citep[CALIFA;][]{califa1} survey of around 600 ETGs, \cite{tsa17} found the volume corrected fraction of prolate rotators of 9\,\%. 

\cite{tsa17} also noticed that the fraction of prolate rotators is much higher among ETGs with stellar masses $M_{*}>2\times10^{11}$\,M$_{\sun}$, around 27\,\% for both CALIFA and ATLAS$^{3{\rm D}}$ surveys.
The fraction is even higher for very massive ETGs. 
\cite{m3g} detected prolate-like rotation in about half of the sample of 25 MUSE Most Massive galaxies (M3G). 
However, these results are not confirmed by the MASSIVE survey of 90 most massive ETGs within a distance of 108\,Mpc \citep{massive1}. 
\cite{massive10} found that only 12\,\% of the MASSIVE galaxies are prolate rotators. 
They also found that kinematic misalignment likely does not depend on stellar mass within the sample.
Nevertheless, they did find a dependence on the environment such that the misalignment is more frequent among galaxies in environments with higher galaxy density. 

On the other side of the galaxy-mass range, spectroscopic measurements of individual red-giant-branch stars reveal prolate rotation in two Local-Group dwarfs -- 
the M31 dwarf spheroidal (dSph) satellite Andromeda\,II \citep{ho12,dp17} 
and the transition-type dwarf Phoenix \citep{kach17}. 

Kinematics and other observed properties of Andromeda\,II were reproduced in the simulations of mergers of disky dwarfs \citep{lo14andii,e15andii,f17andii}.
Generally, binary-merger simulations are known to be able to produce prolate rotation \citep[e.g.,][]{nb03,cox06,hof10}. 
\cite{e15andii} showed that the amount of the prolate rotation in the merger remnants of disky progenitors decreases with the orbital angular momentum of the merger and increases with the disk inclination (with respect to the orbital plane).
Similarly, \cite{tsa17} demonstrated the genesis of prolate rotation in simulations with the disk of one of the merger progenitors perpendicular to the merger orbital plane. 

Prolate rotators have been also found in cosmological zoom simulations \citep{naa14}, the large-scale cosmological hydrodynamical simulation Illustris \citep{illprol,li18,bf19}, and the hydrodynamic cosmological Magneticum Pathfinder simulations \citep{sch18,sch20}.
\cite{sch20} examined radial stellar spin profiles and found that they increase with the radius for all 14 prolate rotators in the Magneticum Pathfinder simulations.
In these simulations, the merger history of galaxies with increasing profiles is dominated by gas-rich major merging. 
\cite{illprol} and \cite{li18} found that Illustris prolate rotators are formed in relatively recent, mostly major, and rather dry mergers. 
\cite{illprol} identified a few examples of other formation channels, but basically, all massive prolate rotators were created in major mergers (at least 1:5) during the last 6\,Gyr of the Illustris simulation. 

The angular momentum of prolate rotation comes mainly from the intrinsic rotation of one (usually the primary) or, sometimes, both merger progenitors \citep{e15andii,illprol,tsa17,li18}. 
The term “prolate” refers to an ellipsoid that is rotationally symmetric around the major axis. 
The analyses of Illustris galaxies show that prolate rotators have prolate triaxial but never oblate shapes \citep{illprol,li18}, while oblate-like rotation occurs in the whole range of shapes \citep{bf19,fb20}.

Based on the Illustris galaxies, \cite{illkdc} predicted that prolate rotators should be associated with signs of the recent galaxy merger more than galaxies hosting kinematically distinct cores (KDCs).
According to \cite{illkdc}, the KDCs in Illustris are also produced in mergers. 
The KDC and prolate rotation can even arise simultaneously during the same merger event.  
However, KDCs can more often have other origins, and if their origin is associated with mergers, the mergers can be minor or ancient.

This work aims to test the predicted association between mergers and prolate rotation among the observed galaxies.
We compare a sample of prolate rotators with a reference sample of MATLAS galaxies.
MATLAS \citep[Mass Assembly of early Type gaLAxies with their fine Structures;][]{duc15,matlas20} is a deep imaging survey conducted at the 3.6\,m Canada-France-Hawaii Telescope (CFHT) using the MegaCam imager.
It consists of 177 nearby ETGs of ATLAS$^{3{\rm D}}$ located in environments with low to medium galaxy density, i.e. it excludes Virgo Cluster members except for the galaxies at the outskirts of the cluster. 
The Next Generation Virgo Cluster Survey \citep[NGVS;][]{ngvs1} is an in-depth survey covering the entire Virgo Cluster area.

We construct a sample of 19 known observed prolate rotators in Sect.\,\ref{sec:sample}. 
In Sect.\,\ref{sec:eval}, we evaluate the presence of morphological signs of galaxy interaction in ground-based deep optical images and find the signs in all galaxies of our sample. 
In Sect.\,\ref{sec:dis}, we compare the findings with the MATLAS sample and show that the association of signs of galaxy interaction and prolate rotation in our sample is of a high statistical significance.
Our work, like many others, demonstrates the benefits of combining deep optical photometry with integral-field spectroscopy for the understanding of the formation of galaxies.

\section{Results} 

\subsection{Sample selection} \label{sec:sample}

We compiled a sample of 19 nearby galaxies with prolate rotation.
Our sample is based on the galaxies listed in \cite{tsa17}, where authors attempted to gather all the prolate rotators and prolate-rotation candidates known at that time.
It consists of prolate rotators from ATLAS$^{3{\rm D}}$, CALIFA, and several additional galaxies drawn from the previously published literature. 
We completed the sample with newly-published prolate rotators.
We did not include the prolate-rotation candidates (i.e. cases where the prolate rotation is not convincingly detected) and galaxies known to have the prolate rotation only in the inner parts, which could be qualified preferably as a large-scale KDC.
For a prolate rotator to be included in our sample, the available images of the galaxy need to be sufficiently deep, so that we can compare the results with the findings of the MATLAS survey. 
Images from the DESI Legacy Imaging Surveys \citep[hereafter the Legacy Surveys][see Sect.\,\ref{sec:eval}]{ls19} and images of a similar or better surface-brightness limit are considered sufficient.

From ATLAS$^{3{\rm D}}$, we selected all the galaxies that have the misalignment between the photometric and kinematic axis greater than $45\degrees$ within the error, according to the SAURON data analyses in \cite{a3d2}: NGC\,1222, NGC\,4261, NGC\,4365, NGC\,4406 (M86), NGC\,5557, and NGC\,5485.
The latter is also a part of the CALIFA sample.
We added NGC\,4486 (M87) that is a part of ATLAS$^{3{\rm D}}$, but it does not exhibit convincing prolate rotation in the SAURON data, possibly due to the presence of a KDC.
However, using the MUSE spectrograph, \cite{em14} revealed NGC\,4486 as a convincing prolate rotator. 

From the CALIFA sample, we included all the galaxies listed in Table~1 in \cite{tsa17} except for UGC\,10695 (a possible candidate for prolate rotation) and NGC\,6338 (exhibits prolate rotation only in the inner parts of the galaxy). 
Galaxies included in our sample from CALIFA are NGC\,647, NGC\,810, NGC\,2484, NGC\,4874, NGC\,5216 (Arp\,104), NGC\,6173, and PGC\,021757 (LSBC\,F560-04).

Five other galaxies are listed as clear cases of prolate rotation in the introduction of \cite{tsa17}.
From those galaxies, we did not include NGC\,5982, because the prolate rotation occurs only in the inner parts; see the velocity field from SAURON \citep{saur3} and OASIS instrument \citep{mcd06}.
The galaxy qualifies instead as an oblate rotator with a KDC.
NGC\,7052 \citep{wa88} is also not included in our sample because the available images are not comparable with images in our comparison sample drawn from the MATLAS survey. 
However, we present this prolate rotator, with informative HST data, in Appendix\,\ref{sec:n7052}.
Galaxies included in our sample from the \cite{tsa17} introduction are  
NGC\,1052 \citep{sg79,di86}, 
NGC\,4589 \citep{wa88,mb89}, and
PGC\,018579 \citep[AM\,0609-331;][]{mm86}. 

The recently-reported prolate rotators are NGC\,7252 and galaxies in the MASSIVE survey. 
NGC\,7252 is revealed as having a clear prolate-rotating component in the stellar velocity field by \cite{wea18}, and we included it in our sample.
The kinematic misalignment angle of MASSIVE galaxies is published in \cite{massive10}.
To select prolate rotators, we applied the abovementioned criterion for the misalignment. 
We only took account of the uncertainties of the kinematic position angles because the errors of photometric position angles are not published in the paper. 
Ten MASSIVE galaxies satisfy the criterion. 
Two of them (NGC\,4874 and NGC\,5557) are already included in our sample. 
Three have Legacy Surveys images, but the images of NGC\,2258 and NGC\,2832 are spoiled with artifacts, so we cannot confidently confirm the presence of tidal structures in these galaxies. 
The remaining five do not have sufficiently deep images. 
The only additional prolate rotator from the MASSIVE survey included in our sample is NGC\,2783.

Besides the prolate-rotation candidates, the galaxies with only inner prolate rotation, and prolate rotators without sufficient images, we did not include the two Local-Group dwarfs with prolate rotation because they are far out of the range of galaxy masses of the reference sample. 
At the opposite end, the M3G survey targets the most massive galaxies in the densest galaxy environments at $z\approx0.045$ \citep{m3g}. Therefore, we did not include them either.
The final sample of 19 prolate rotators is listed in Table\,\ref{tab:prs}.

\begin{table*}
\caption{The sample of prolate rotators}
\label{tab:prs}
\centering
\begin{tabular}{cccccccc} 

\hline\hline 
(1) & (2) & (3) & (4) & (5) & (6) & (7) \\
Name & Other names & Survey & $\Psi$ & $D$ & $M_{K}$ & log($M_{\rm JAM}$) \\
 & & & (deg) & (Mpc) & (mag) & \\
\hline 
NGC\,1222 & & A3D & $ 73 \pm 15 $ \tablefootmark{a} & 33.3 \tablefootmark{f} & -22.71 & 10.50 \tablefootmark{k} \\
NGC\,4261 & & A3D & $ 74 \pm 3 $ \tablefootmark{a} & 30.8 \tablefootmark{f} & -25.19 & 11.72 \tablefootmark{k} \\
NGC\,4365 & VCC\,731 & A3D & $ 76 \pm 7 $ \tablefootmark{a} & 23.3 \tablefootmark{f} & -25.20 & 11.53 \tablefootmark{k} \\
NGC\,4406 & M86 & A3D & $ 81 \pm 13 $ \tablefootmark{a} & 16.8 \tablefootmark{f} & -25.03 & 11.60 \tablefootmark{k} \\
NGC\,4486 & M87 & A3D & $ 46 \pm 58 $ \tablefootmark{a} & 17.2 \tablefootmark{f} & -25.37 & 11.73 \tablefootmark{k} \\
NGC\,5557 & & A3D, MAS & $ 73 \pm 6 $ \tablefootmark{a} & 38.8 \tablefootmark{f} & -24.87 & 11.33 \tablefootmark{k} \\
NGC\,5485 & & A3D, CAL & $ 80 \pm 3 $ \tablefootmark{b} & 25.2 \tablefootmark{f} & -23.61 & 11.06 \tablefootmark{k} \\
NGC\,0647 & & CAL & $ 72 \pm 3 $ \tablefootmark{b} & 184 \tablefootmark{b} & - & ($\sim 11.57 $) \tablefootmark{b} \\
NGC\,0810 & & CAL & $ 87 \pm 2 $ \tablefootmark{b} & 110 \tablefootmark{b} & -25.67 & $\sim 11.7 $ \\
NGC\,2484 & & CAL & $ 52 \pm 4 $ \tablefootmark{b} & 192 \tablefootmark{b} & -26.56 & $\sim 12.1 $ \\
NGC\,4874 & & CAL, MAS & $ 86 \pm 5 $ \tablefootmark{b} & 102 \tablefootmark{b} & -26.19 & $\sim 12.0 $ \\
NGC\,5216 & Arp\,104 & CAL & $ 66 \pm 5 $ \tablefootmark{b} & 42 \tablefootmark{b} & -23.25 & $\sim 10.7 $ \\
NGC\,6173 & & CAL & $ 80 \pm 2 $ \tablefootmark{b} & 126 \tablefootmark{b} & -26.14 & $\sim 11.9 $ \\
PGC\,021757 & LSBC\,F560-04 & CAL & $ 57 \pm 5 $ \tablefootmark{b} & 238 \tablefootmark{b} & -26.79 & $\sim 12.2 $ \\
NGC\,1052 & & - & $\sim90$ \tablefootmark{c} & 19.4 \tablefootmark{g} & -24.00 & $\sim 11.0 $ \\
NGC\,2783 & UGC\,04859 & MAS & $ 76 \pm 11 $ \tablefootmark{d} & 85.8 \tablefootmark{h} & -25.35 & $\sim 11.6 $ \\
NGC\,4589 & & - & $\sim45$ \tablefootmark{e} & 22.0 \tablefootmark{g} & -23.96 & $\sim 11.0 $ \\
NGC\,7252 & Arp\,226 & - & - & 60.1 \tablefootmark{i} & -24.59 & $\sim 11.3 $ \\
PGC\,018579 & AM\,0609-331 & - & - & 129 \tablefootmark{j} & -24.48 & $\sim 11.2 $ \\
\hline 
\end{tabular}
\tablefoot{
(1)	the galaxy name in the NGC or PGC catalog; 
(2)	an alternative galaxy name; 
(3)	A3D, CAL, MAS -- the galaxy is a part of the ATLAS$^{3{\rm D}}$, CALIFA, or MASSIVE survey, respectively; 
(4)	 $\Psi$, the misalignment angle between the kinematic and photometric axis, where available; 
(5)	 $D$, the distance of the galaxy; 
(6)	 $M_{K}$, total galaxy absolute magnitude, see Sect.\,\ref{sec:mass}; 
(7)	 log($M_{\rm JAM}$), the dynamical mass of the galaxy -- the ATLAS$^{3{\rm D}}$ value is adopted where possible, the rest are approximate estimates calculated from a $M_{K}$\,--\,log($M_{\rm JAM}$) relation, except for NGC\,0647, where the stellar mass estimate from \cite{tsa17} is listed instead, see Sect.\,\ref{sec:mass}.\\
\tablefoottext{a}{ATLAS$^{3{\rm D}}$ \citep{a3d2},}
\tablefoottext{b}{CALIFA \citep{tsa17},}
\tablefoottext{c}{\cite{di86},} 
\tablefoottext{d}{MASSIVE \citep{massive10},}
\tablefoottext{e}{\cite{mb89},}
\tablefoottext{f}{ATLAS$^{3{\rm D}}$ \citep{a3d1},}
\tablefoottext{g}{\cite{sfb01},}
\tablefoottext{h}{MASSIVE \citep{massive1},}
\tablefoottext{i}{\cite{dist07},}
\tablefoottext{j}{\cite{dist13},}
\tablefoottext{k}{ATLAS$^{3{\rm D}}$ \citep{a3d15}.}
}
\end{table*}

\subsection{Sample masses} \label{sec:mass}

In this paper, we compare the incidence of the morphological signs of galaxy interactions in these prolate rotators with the incidence in the MATLAS sample. 
Since the incidence shows a significant dependence on the host galaxy mass, it is important to have a general idea of the mass distribution of galaxies in our sample.
The masses of the MATLAS galaxies are expressed as the dynamical mass, log($M_{\rm JAM}$), obtained by Jeans Anisotropic Modeling \citep{cap08} within the half-light radius.
The values were derived in \cite{a3d15} from the observed kinematic maps of the galaxies for the whole ATLAS$^{3{\rm D}}$ sample.
 
For the prolate rotators outside the ATLAS$^{3{\rm D}}$ sample, we derive approximate estimates from the total galaxy absolute magnitude, $M_{K}$ (Column (6) in Table\,\ref{tab:prs}).
The value of $M_{K}$ is computed in the same way as it was done for the ATLAS$^{3{\rm D}}$ sample in \cite{a3d1}.
It is derived from the apparent magnitude $K_T$ (the keyword \texttt{k\_m\_ext}) from the Two Micron All Sky Survey (2MASS) extended source catalog \citep[XSC;][]{2massesc}. 
Then $M_{K}$ is calculated as $M_K=K_T-5\log_{10} D - 25 - A_B/11.8$, where $D$ is the distance of the galaxy in Mpc (Column (5) in Table\,\ref{tab:prs}). $A_B$ values for the correction for the foreground galactic extinction are adopted from \cite{ext11}. 

There is a correlation between log($M_{\rm JAM}$) and $M_{K}$.
We fitted the available data of 258 ATLAS$^{3{\rm D}}$ galaxies with linear regression. 
The RMS of residuals of the fit is 0.15.
We used the fitted function to calculate approximate values of log($M_{\rm JAM}$) for the rest of the sample (Column (7) in Table\,\ref{tab:prs}). 
This calculation method was not possible only for NGC\,647 since the galaxy lacks the \texttt{k\_m\_ext} value in XSC. 
For this galaxy, we listed the stellar mass estimate from \cite{tsa17} instead.

We do not want to compare the exact mass distributions of the galaxies.
For our purposes, see Sect.\,\ref{sec:sig}, it is sufficient to have the general idea that almost all of our prolate rotators have log($M_{\rm JAM}$), or its approximate equivalent, greater than 11 with the highest values around 12. 
Given the precision of our estimates, the maximum is roughly comparable with the MATLAS survey.

\begin{figure*} 
\centering
\includegraphics[width=\hsize]{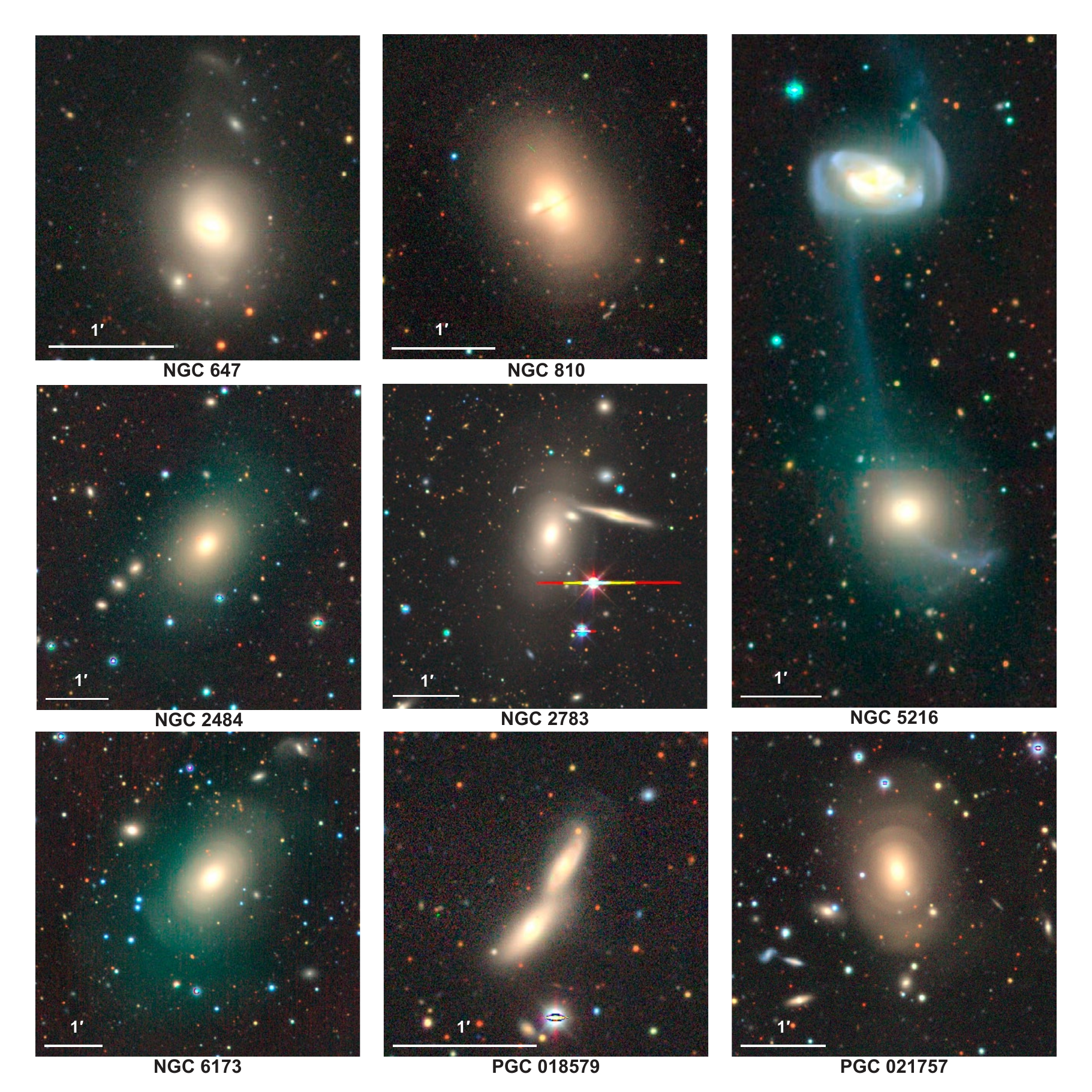}
\caption{
Images of eight prolate rotators from our sample extracted from the Legacy Surveys. 
North is up, east is to the left.
\label{fig:ls7}
}
\end{figure*}

\begin{figure} 
\centering
\includegraphics[width=\hsize]{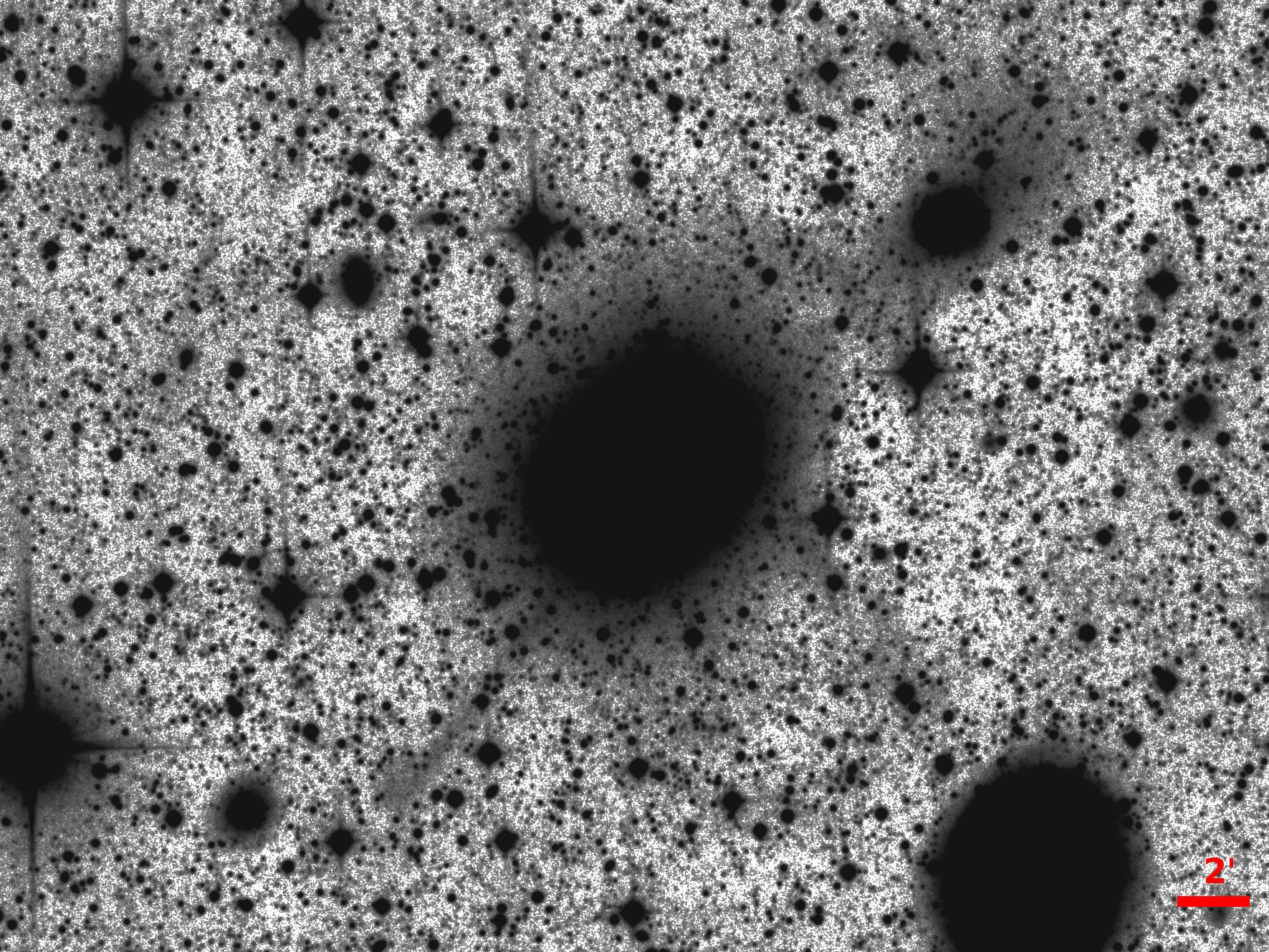}
\caption{
Negative image of NGC\,1052 from the HERON survey \citep{heron19}, data processed by Javier Rom{\'a}n and Oliver M{\"u}ller. 
\label{fig:1052}
}
\end{figure}

\begin{figure*} 
\centering
\includegraphics[width=0.95\hsize]{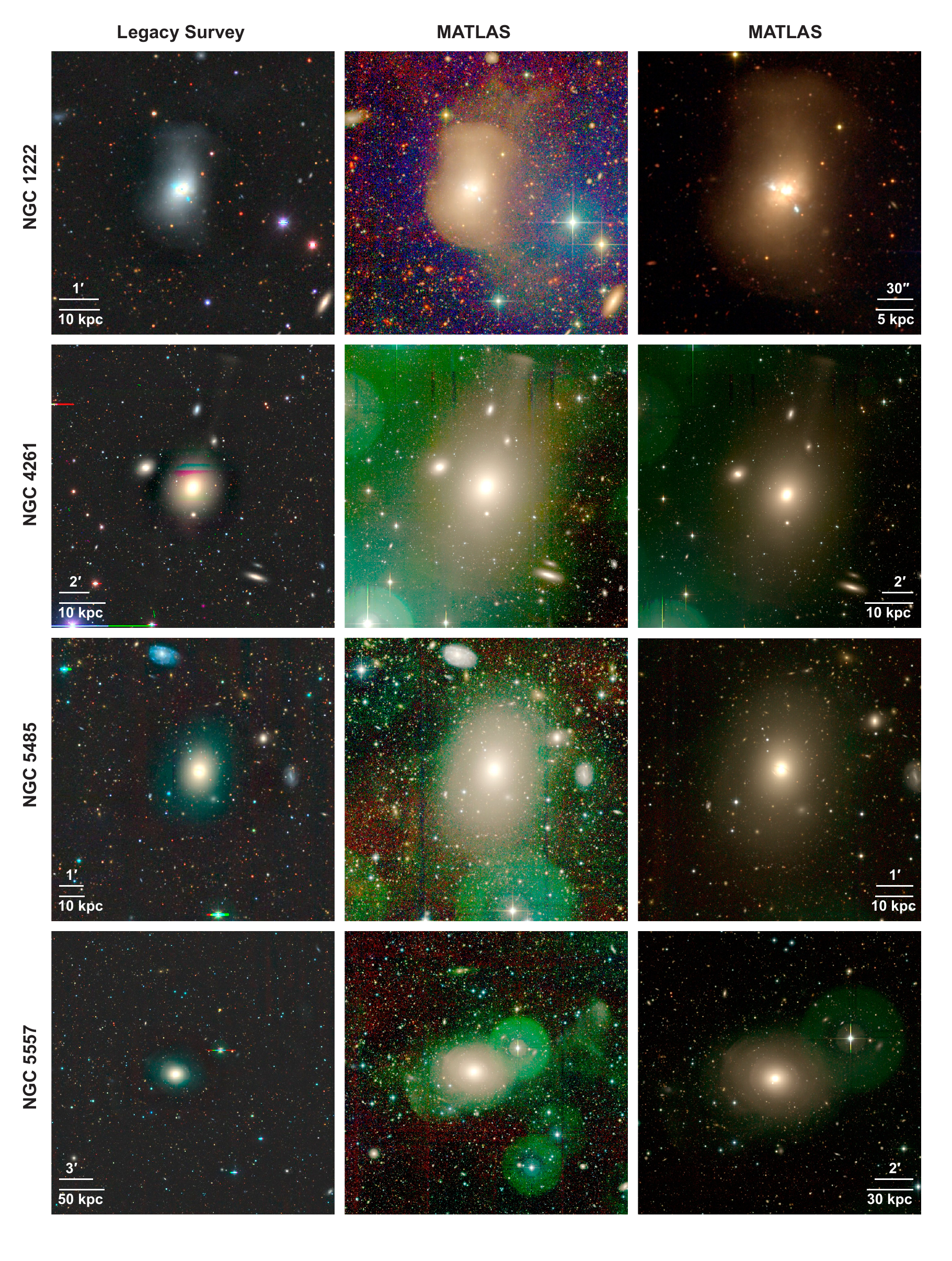}
\caption{
Images of four prolate rotators from our sample. 
Left column: image extracted from the Legacy Surveys.
Middle column: image from MATLAS showing the outer parts of the stellar halo with the same field of view as the image in the left column of the respective row.
Right column: image from MATLAS showing more inner parts of the galaxy. 
North is up, east is to the left.
\label{fig:lsma}
}
\end{figure*}

\begin{figure} 
\centering
\includegraphics[width=\hsize]{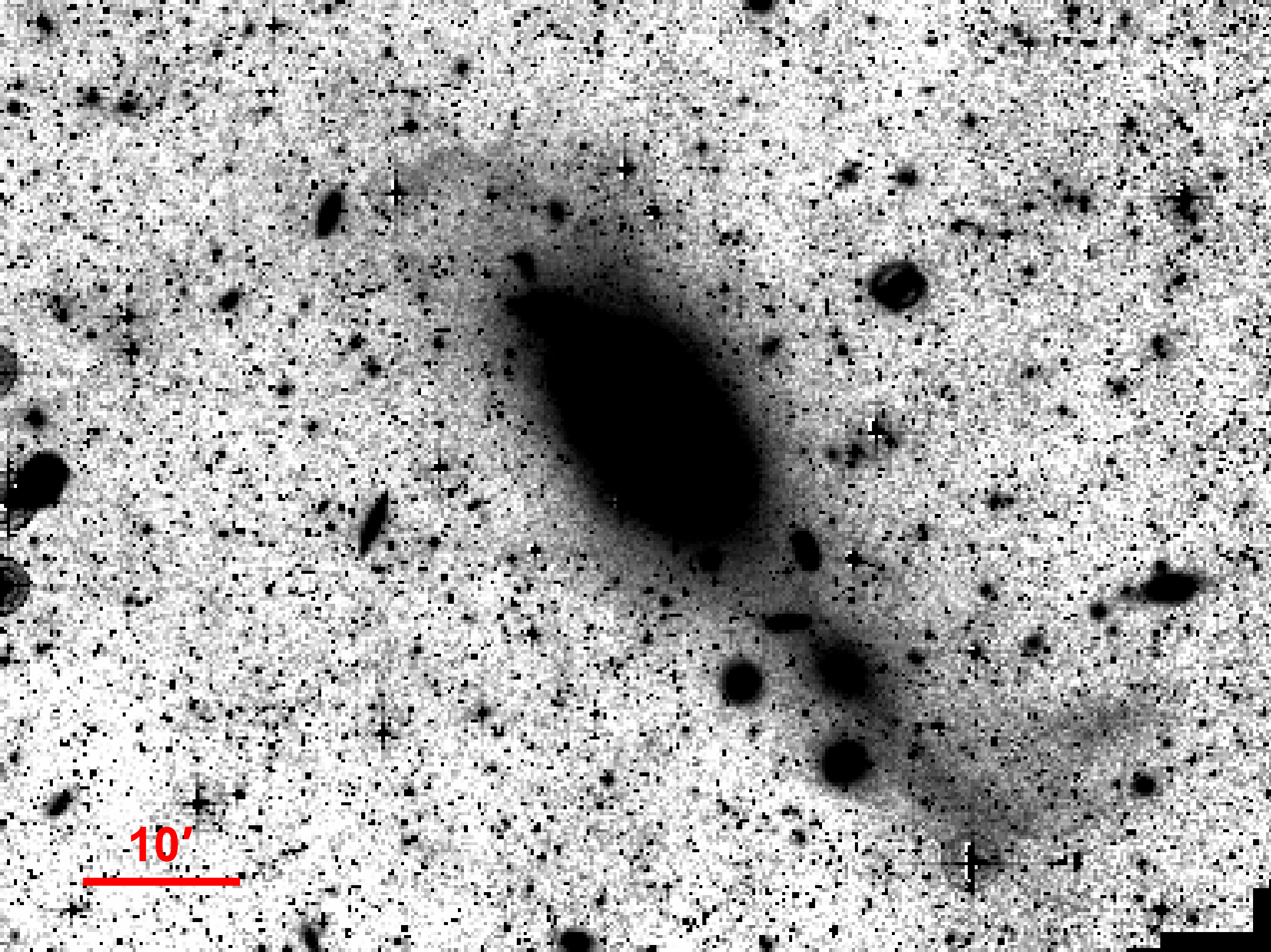}
\caption{
Negative image of NGC\,4365 from the Burrell Schmidt telescope.
North is up, east is to the left. 
Image credit: The Burrell Schmidt Deep Virgo Survey, \cite{mih17}.
\label{fig:4365}
}
\end{figure}

\begin{figure*} 
\centering
\includegraphics[width=\hsize]{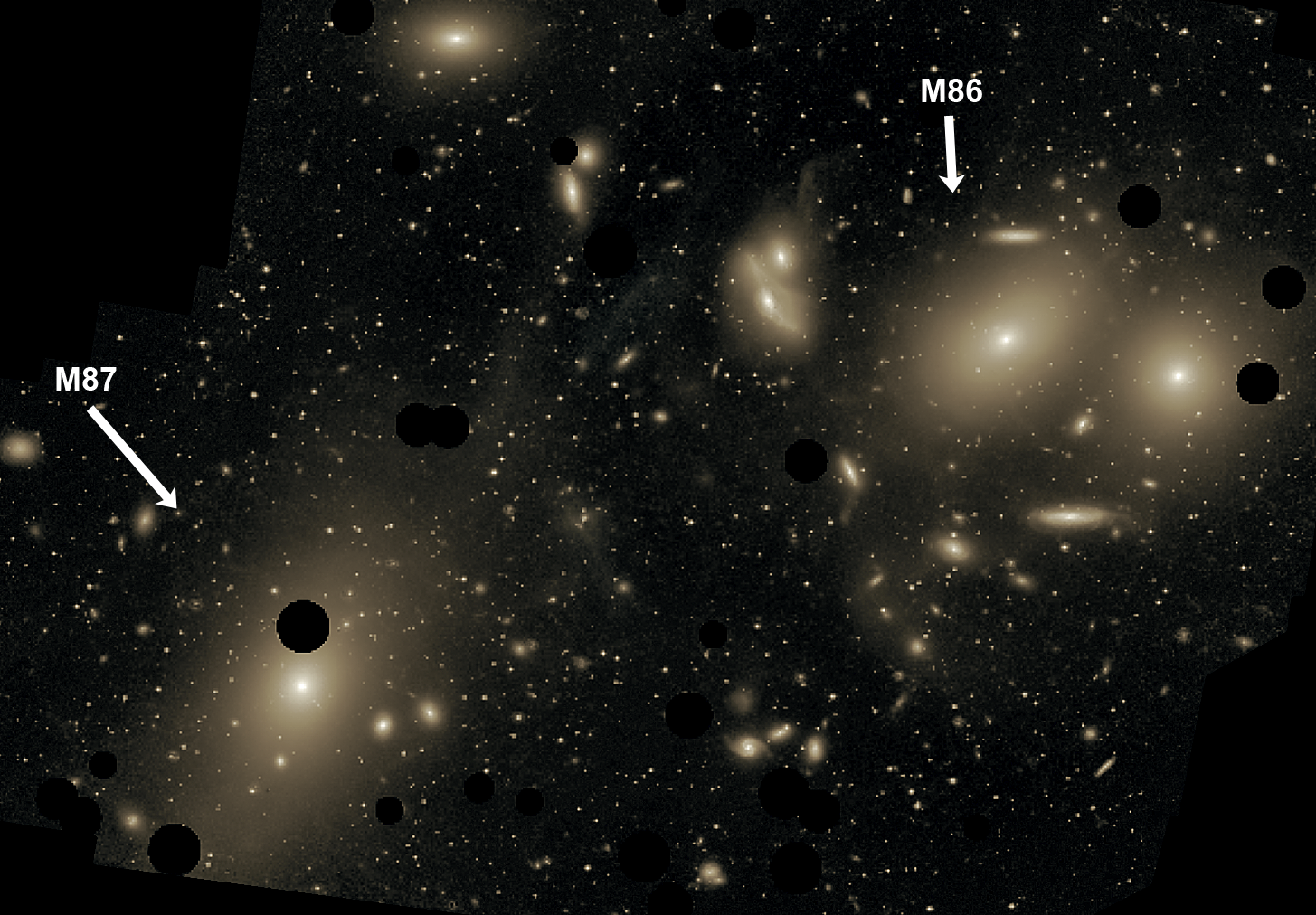}
\caption{
Images of two prolate rotators from our sample. 
The deep image of the Virgo Cluster was obtained by Chris Mihos and his colleagues using the Burrell Schmidt telescope.
The two prolate rotators from our sample (M86 and M87) are the two biggest galaxies in the image.
The dark spots indicate where bright foreground stars were removed from the image.
North is up, east is to the left. 
Image credit: Chris Mihos (Case Western Reserve University)/ESO.
\label{fig:virgo}
}
\end{figure*}

\begin{figure*} 
\centering
\includegraphics[width=\hsize]{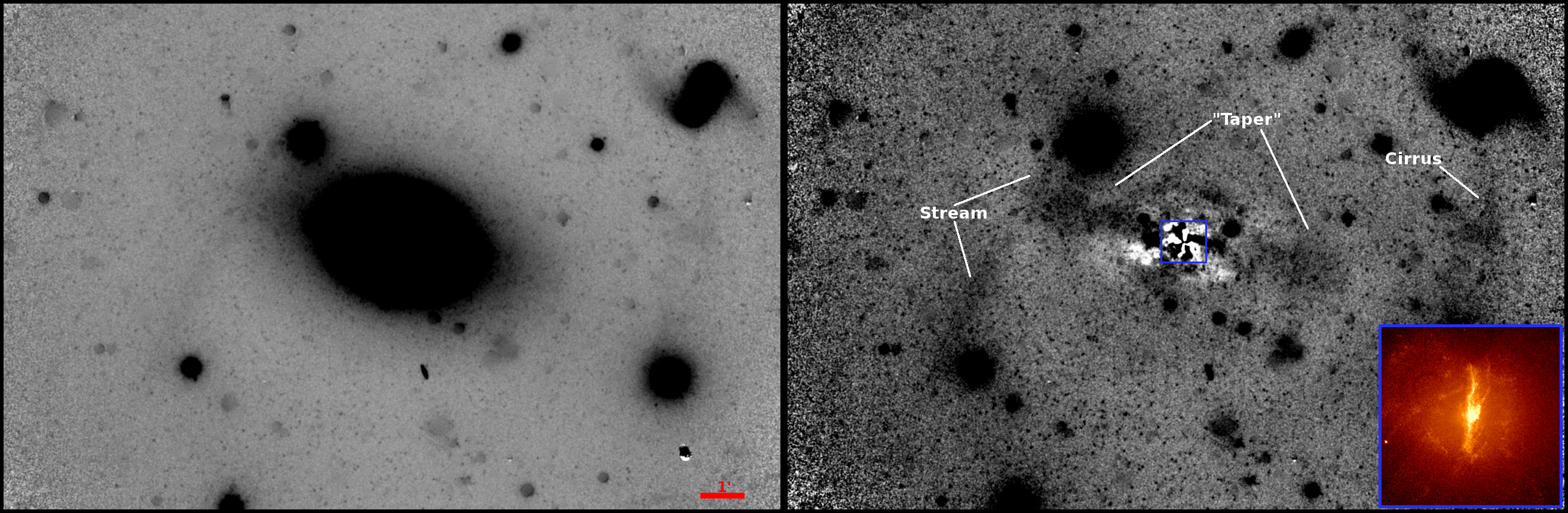}
\caption{
Negative images of NGC\,4589. Left: the deep optical image of the galaxy by a 1.4-meter Milankovi\'c telescope; right: the residual image of the one in the left panel. Both panels have the same field of view. The image in the left panel has been cleaned (see the text for the details) and the small disk-like shapes are remnants of this cleaning process. 
The inset shows a closer view of the central region indicated by a blue box in the right panel. It is an HST color image obtained by using the F814W-F435W filters. The dust filaments, especially the one perpendicular to the major axis, are apparent in the HST data. North is up, east is to the left.
\label{fig:4589}
}
\end{figure*}

\begin{figure*} 
\centering
\includegraphics[width=\hsize]{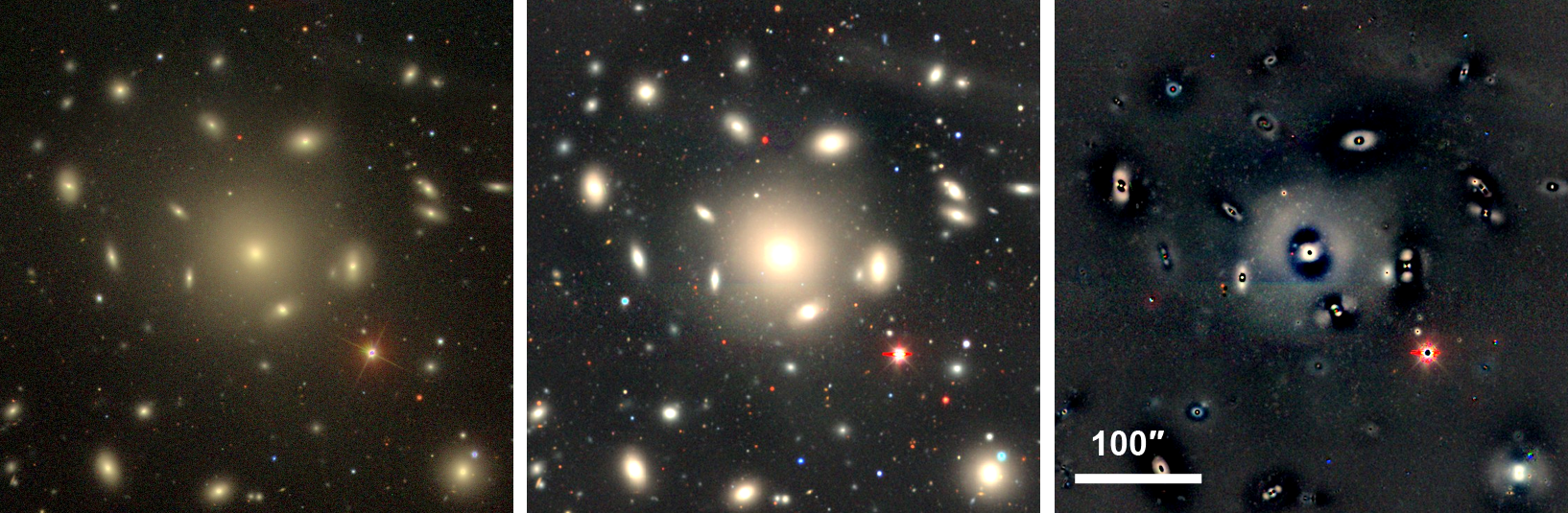}
\caption{
Images of NGC\,4874, one of the prolate rotators from our sample. 
Left: the SDSS image; middle: the Legacy Surveys image; right: the Legacy Surveys residual image. 
All panels have the same field of view.
North is up, east is to the left.
\label{fig:4874}
}
\end{figure*}

\begin{figure*} 
\centering
\includegraphics[width=\hsize]{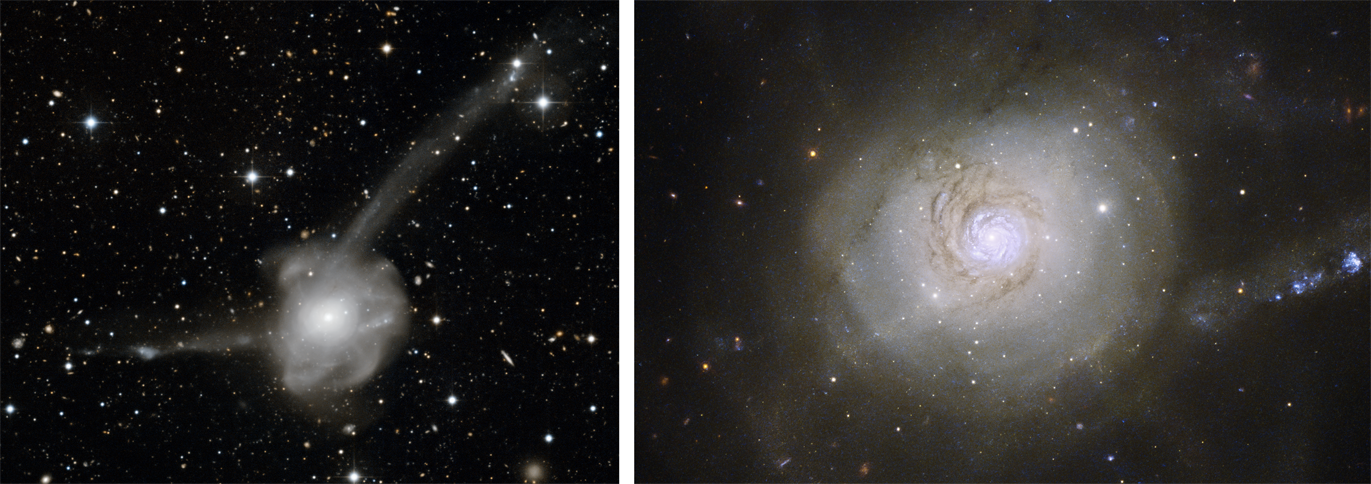}
\caption{
Images of NGC\,7252, one of the prolate rotators from our sample. 
Left: Image taken by the Wide Field Imager on the MPG/ESO 2.2-metre telescope at ESO’s La Silla Observatory in Chile with a total exposure time of more than four hours; Image credit: ESO.
Right: HST/WFC3 image of the inner parts of the galaxy processed by Judy Schmidt to emphasize fine structures; Image credit: NASA \& ESA, Acknowledgement: Judy Schmidt. 
North is up, east is to the left.
\label{fig:7252}
}
\end{figure*}

\subsection{Ubiquitous interaction signs in the sample} \label{sec:eval}

Here, we evaluate the presence of morphological signs of galaxy interactions in the prolate rotators in our sample. 
In Sect.\,\ref{sec:sig}, we compare the frequency of the tidal disturbance in prolate rotators with the frequency reported for the MATLAS galaxies \citep{matlas20}. 
Images of the prolate rotators, evaluated in this section, come from several sources, but in Sect.\,\ref{sec:sbl}, we establish that all these images have surface-brightness limits similar or worse than the MATLAS images.

In compliance with the MATLAS classification, the signs of galaxy interaction are tidal features (tails, plumes, loops, streams, and stellar shells) and disturbed or asymmetric outer isophotes of the stellar halo. 
Tails or plumes usually come from a major merger. They are thick, elongated features directly attached to the body of the galaxy in question. 
Streams are usually thinner, sometimes detached, features typically created by stripping a lower-mass companion galaxy.
Shells, also called ripples, are azimuthal arcs centered on the core of the host galaxy, usually with sharp outer edges.
 
Shell galaxies have been believed to result from close-to-radial minor mergers for decades
\citep[e.g.,][]{q84, dc86, hq88, e12sg}. However, the view concerning their origin has recently shifted from minor mergers to intermediate-mass \citep{duc15,e20sg} or even major ones \citep{illsg17}.
Three morphological types of shell systems are recognized \citep{wil87c,pri90,wil90,wil00}:
\begin{itemize}
\setlength\itemsep{0.3em}
\renewcommand{\labelitemi}{$\bullet$}
\item Type\,I (“Cone” or “Aligned”)\,--\,typically found in galaxies with a higher ellipticity. Shells are interleaved in radius on alternate sides of the galaxy, i.e. the next outermost shell usually lies on the opposite side of the galaxy center. 
They are well-aligned with the major photometric axis of the galaxy, mostly confined in a bi-conical structure. 
Shell separation increases with radius. 
Examples of such shells are NGC\,810, 2484, 2783, 6173, and PGC\,021757, all shown in Fig.\,\ref{fig:ls7}.
\item Type\,II (“Randomly distributed arcs” or “All round”)\,--\,the position angles of the shells are randomly distributed all around a somewhat circular galaxy. 
For example, NGC\,5216 (Fig.\,\ref{fig:ls7}) or the famous NGC\,474 (not a prolate rotator).
\item Type\,III (“Irregular”)\,--\,shell systems with more complex structures or have too
few shells to be classified.
\end{itemize}
In the catalog of 137 shell galaxies \cite{mc83}, all three types occur in approximately the same fraction \citep{pri90}. 
Shell systems are appealing because they can be, in principle, used to constrain the time of the merger and the gravitational potential of the host galaxy \citep{q84,dc86,hq87a,hq87b,mk98,can07,sh13,bil13,bil14,bil15,e12sg,e19sgIAU,e20sg}.

To inspect our sample, we used archival images from the
MATLAS survey \citep[][\footnote{\url{https://obas-matlas.u-strasbg.fr}}]{duc15,matlas20}, 
the DESI Legacy Imaging Surveys \citep[the Legacy Surveys;][\footnote{\url{https://www.legacysurvey.org}}]{ls19}, 
the Sloan Digital Sky Survey (SDSS),  
the Burrell Schmidt Deep Virgo Survey \citep[][\footnote{\url{http://astroweb.case.edu/VirgoSurvey/}}]{mih17}, 
the Halos and Environments of Nearby Galaxies (HERON) survey \citep{heron19}, 
and an image from Wide Field Imager on the MPG/ESO 2.2-metre telescope at ESO’s La Silla Observatory.
One of the galaxies (NGC\,4589) was examined in images recently obtained with the 1.4-meter Milankovi\'c telescope mounted at the Astronomical Station Vidojevica in Serbia.  
We describe the details of the observation in Appendix\,\ref{apx:obs}.
In several cases, we also show complementary images from the Hubble Space Telescope (HST), but we never use them as a source to evaluate the tidal disturbance of the galaxies in our sample.

Our findings are summarized below for each galaxy separately. 
We also added notes to individual galaxies about the presence of other features, like KDCs and dust lanes, or other relevant information from the literature.

\subsection*{NGC\,647} 

NGC\,647 has clear interaction signs, most notably a distinct plume or loop stretching around 80\arcsec{} to the north, visible in the Legacy Surveys image, Fig.\,\ref{fig:ls7}. 
The galaxy also displays a dust line in the center, roughly allied with the minor photometric axis.
Two close smaller galaxies (one 50\arcsec{} to the north, the other 25\arcsec{} to the southeast) are candidates for the surviving cores of the merging secondary galaxies.

\subsection*{NGC\,810} 

NGC\,810 is an excellent example of a rich Type\,I shell system with at least eight shells detected in the Legacy Surveys image, Fig.\,\ref{fig:ls7}. 
The galaxy also possesses a prominent dust lane along the minor photometric axis.
Around 10\arcsec{} from the center, close to the dust line, there is a possible surviving core of the merging secondary galaxy.

\subsection*{NGC\,1052} 

NGC\,1052, Fig.\,\ref{fig:1052}, is a dominant member of a galaxy group that became popular recently because it hosts two supposedly dark-matter-free ultra-diffuse galaxies, NGC\,1052-DF2 and NGC\,1052-DF4 \citep{vand18,vand19}. 

The numerous signs of interactions in NGC\,1052 have been summarized in \cite{mu19}. 
Their deep optical image \citep[Figure\,1 in][]{mu19} uncovers a 10\arcmin{} long stream pointing southeast, a loop in the southwest, and a shell-like structure in the northwest between NGC\,1052 and NGC\,1047.
Moreover, an H\,I feature in the galaxy center seems to be associated with the optical tidal bridge leading to NGC\,1047. 
The galaxy hosts a faint diffuse kpc-scale dust lane situated roughly along the minor axis \citep{ca83,di86}. 
The star and gas kinematics are largely (about 66\degrees) misaligned \citep{di86}. 
Finally, the modeling suggests a starburst in the center 1\,Gyr ago due to a recent merger with a gas-rich galaxy \citep{vang86,pie05}. 

\subsection*{NGC\,1222} 

NGC\,1222, Fig.\,\ref{fig:lsma}, has around six shells visible in the MATLAS image. 
The Type\,I shell system is somewhat atypical, with more sharp shells on one (southern) side. 
There are additional asymmetric faint tidal features in the galaxy outskirts (both north and south). A significant amount of atomic hydrogen was detected in the galaxy, and the gas and stellar kinematics are roughly aligned \citep{yng18}.
Abundant irregularly shaped dust in the inner region tends to lay more along the minor photometric axis.
In the MATLAS survey, NGC\,1222 is classified as having shells, tails, disturbed outer isophotes, and prominent dust lanes \citep{matlas20}.

\subsection*{NGC\,2484} 

NGC\,2484 is a BCG (brightest cluster galaxy) hosting a nice Type\,I shell system with at least four shells detectable in the Legacy Surveys image, Fig.\,\ref{fig:ls7}.

\subsection*{NGC\,2783} 

NGC\,2783 is the brightest member of a small compact group of five galaxies, Hickson\,37 \citep{rub91,wch95}.
It is another rich Type\,I shell system with at least six shells visible in the Legacy Surveys image, Fig.\,\ref{fig:ls7}. 
The galaxy also contains a rapidly rotating disk of ionized gas \citep{cal89}.

\subsection*{NGC\,4261} 

NGC\,4261, Fig.\,\ref{fig:lsma}, is a member of the Virgo cluster located in the outskirts of the cluster.
The MATLAS image reveals a narrow shell or umbrella feature stretching around 9\arcmin{} (29\,kpc) to the north and a probable fainter but wider shell closer to the center (around 7\arcmin, 22\,kpc) on the other side of the galaxy.
The galaxy also has dust lanes situated roughly along the major axis \citep{mb87,mah96} and a disk of cool dust and gas surrounding the nucleus \citep{jaf93}. 
In the MATLAS survey, NGC\,4261 is classified as having shells and disturbed outer isophotes \citep{matlas20}. 

\subsection*{NGC\,4365 (VCC\,731)} 

NGC\,4365, Fig.\,\ref{fig:4365}, is the dominant galaxy of the Virgo W$^\prime$ group, located $\sim6$\,Mpc behind the Virgo Cluster \citep{mei07}. 
The galaxy is long known to have a KDC \citep{be88,bs92,bsg94,for95} and has been extensively studied also through the integral field spectroscopy (SAURON\,--\,\citealt{dav01} and \citealt{a3d2} and MUSE\,--\,\citealt{ned19}). 

Several substantial tidal features have been described, see Figure\,6 in \cite{bog12} and Figure\,9 in \cite{mih17}: 
an asymmetric S Plume; 
NE Loop around 26\arcmin{} (146\,kpc) from the center; 
and most notably, a long SW Tail with a diffuse shell-like structure at its ending, around 40\arcmin{} (230\,kpc) from the center. 
A companion galaxy, NGC\,4342, sits in the middle of the tail.
\cite{blom12a,blom12b,blom14} studied globular clusters in the system.  
They found an overdensity of globular clusters in the tail. 
Their analysis of the colors and kinematics of the globular clusters suggests that they have been tidally stripped from NGC\,4342. 
A similar conclusion was reached by \cite{mih17}, analyzing the colors and morphology of the diffuse stellar light of the tail.
Moreover, NGC\,4342 itself is an outlier on galaxy scaling relations in a way that is consistent with the galaxy being tidally stripped \citep{blom14}.

\subsection*{NGC\,4406 (M86)} 

M86 is a massive elliptical galaxy in the Virgo cluster. 
It is long known to have a KDC \citep{be88,bs92,bsg94} and has also been studied by integral field spectroscopy \citep{a3d2}. 

M86 is a known shell galaxy \citep{for94}. A deep image reveals around five shells in the outskirts, see Fig.\,\ref{fig:virgo}.
The stellar halo hosts several other substructures like several low-surface-brightness filaments, a tidal bridge, stream, and plumes \citep{mih05,mih17,jnw10}. 
Some substructures cannot be safely assigned to M86 since its halo is partially overlapped, in projection, with the halo of M84, although the galaxies themselves may be well separated along the line of sight \citep{mei07}. 

\subsection*{NGC\,4486 (M87)} 

M87 is the central BCG of the Virgo cluster. 
It does not show convincing prolate rotation in the SAURON data \citep{a3d2}. However, the MUSE spectrograph fully reveals the prolate character of the large-scale rotation as well as the presence of a KDC \citep{em14}. 

The extended stellar halo hosts a number of streams (often associated with satellite galaxies in the process of tidal striping) as well as signatures of more massive mergers, as summarized in \cite{mih17}. 
M87 is long known to have asymmetric outer isophotes \citep{cd78}. 
A long (38\arcmin, 178\,kpc) NW stream along the major photometric axis connects the stellar halo to the pair of galaxies NGC\,4458 and NGC\,4461 \citep{mih05,jnw10,rud10}. 
In the same direction but closer to the galaxy (20\arcmin, 90\,kpc), there is a primarily discussed shell-like feature, the so-called “cap” or “crown” \citep{weil97,lon15b,lon15c}. 
It may be connected to a kinematic substructure discovered in planetary nebulae by  \cite{lon15b}. 
There is also a kinematic substructure in globular clusters reported by \cite{rom12}. 
The kinematic substructures suggest a past merger event with a secondary galaxy of a mass of about $10^{10}$\,M$_{\sun}$.

Besides the crown, there are hints of other diffuse shell-like structures in the direction of the major photometric axis, giving the galaxy an appearance of an almost fade-out Type\,I shell system, see Fig.\,\ref{fig:virgo}.

\subsection*{NGC\,4589} 

NGC\,4589 is part of a small group.
The galaxy also possesses a KDC \citep{for95}.
It was not possible to assess the galaxy from the Legacy Surveys images due to the abundance of image artifacts.
The left panel of Fig.\,\ref{fig:4589} shows a new deep optical image of NGC\,4589 obtained with the 1.4-meter Milankovi\'c telescope. For more details, see Appendix\,\ref{apx:obs}. 
All the features described below are visible without a model subtraction, but we show the residual image and emphasize the detected substructures.

We cut the original image to remove the noisy edges and smoothed it with a Gaussian function to bring out the faint structures. 
The right panel of Fig.\,\ref{fig:4589} shows the smoothed residual image of NGC\,4589.

Having cleared and masked the original image, we model the galaxy by using the \textsc{iraf}'s \textsc{ellipse} routine. The position angle and central coordinate parameters are set free during the ellipse fitting. We subtract the model from the cleaned image of NGC\,4589 to obtain the residual image. To reveal the faint structures, we smooth this residual image with the same Gaussian function used for the original image. 
The cleaning process masks small and faint objects in the image. We do not take into account these masked regions during the ellipse fitting routine. After the ellipse fitting, we fill these masked regions with random pixel values drawn from a Gaussian distribution for cosmetic reasons. We determine the Gaussian distribution from a limited area surrounding the faint and small objects. However, the resulting mask seems like a disk-like artifact if the surrounding area contains a bright pixel. 

There is a cirrus (a foreground dust cloud situated in our Galaxy) 7\arcmin{} to the west from the galaxy center (near the right side of the images in Fig.\,\ref{fig:4589}). 
It is at the border of a group of other cirri (outside the image), showing a typical morphology with filamentary substructures \citep{duc15}, similar to other cirri in the area, and all these structures are visible in the WISE 12-micron dust map. 
We do not consider this feature to be a sign of galaxy interaction. 

On the east side, 3.6\arcmin{} from the center, there is a linear structure nearly perpendicular to the major photometric axis of the galaxy.
It is marked as a “stream” in the right panel of Fig.\,\ref{fig:4589}.
It does not look like a cirrus -- it is far from the group of the cirri and shows no substructure. 
Moreover, unlike the cirrus in the west, this structure is not visible in the WISE 12-micron dust map, which is incorporated in the Legacy Surveys viewer.
At the extension of the stream in the Legacy Surveys image, 11\arcmin{} to the southeast, there is a faint galaxy, probably a dwarf (assuming the same distance as NGC\,4589), a possible progenitor of the stream.

The most notable feature in NGC\,4589 is an elongated asymmetric excess of light roughly along the major photometric axis. 
We call it a “Taper” (see the right panel of Fig.\,\ref{fig:4589}). 
It is about 8\arcmin{} long, thicker on the west side, and piercing through the stream on the east side. 
Such an irregular structure could be a tidal tail or a disk disrupted by a past galaxy interaction. 

A companion galaxy NGC\,4572, the northwest corner of Fig.\,\ref{fig:4589}, is tidally disturbed. 
The deep image reveals prominent tidal arms. 
\cite{s4g15} considered NGC\,4572 to be a strong candidate for a significantly warped disk with much extraplanar material or a strong tidal spiral.
NGC\,4589 the only bigger galaxy in its surroundings. 
The radial-velocity difference of the galaxies, 230\,km\,s$^{-1}$ \citep[SIMBAD,][]{simbad}, is consistent with them being associated.

The galaxy has a strong dust lane parallel to the minor photometric axis \citep{mb89,gou94,wik95,car97}, also imaged by HST, see the inset in the right panel of Fig.\,\ref{fig:4589}.
Studying gas and stellar kinematics and dust morphology, several authors concluded that this galaxy is a merger remnant \citep{mb89,rav01,hak08}.

\subsection*{NGC\,4874} 

NGC\,4874 has weak but clear prolate rotation. 
It is one of two dominant members of the Coma cluster (Abell\,1656), and it is considered a BCG \citep[e.g.,][]{liu10}. 

The stellar halo of the galaxy is highly asymmetrical, lopsided towards the northeast direction. 
This asymmetry is not apparent in the Legacy Surveys image, the middle panel of Fig.\,\ref{fig:4874}, due to several artifacts and the local background subtraction. 
Fortunately, the lopsided halo becomes evident in the SDSS image, the left panel of Fig.\,\ref{fig:4874}.
The galaxy also possesses faint shells that are barely visible in the Legacy Surveys image, but they are present when looking at the residuals available in the Legacy Surveys viewer, the right panel of Fig.\,\ref{fig:4874}.

\subsection*{NGC\,5216 (Arp\,104)} 

NGC\,5216 is a Type\,II shell galaxy with at least four distinct shells visible in the Legacy Surveys image, Fig.\,\ref{fig:ls7}.
There is an on-going mass transfer from a bluer companion galaxy -- a thin blue tail stretches across the whole body of NGC\,5216 and extends about 3\arcmin{} north, connecting to the strongly disturbed NGC\,5218.
The shells of NGC\,5216 can originate from a previous interaction with the companion galaxy NGC\,5218 or another past merger event.

\subsection*{NGC\,5485} 

In the MATLAS image, Fig.\,\ref{fig:lsma}, NGC\,5485 has a disturbed asymmetrical stellar halo with a  stream or tail pointing towards a possible faint shell in the southeast.
A deep color image of NGC\,5485 shows a ring-like dust structure in the central region oriented along the minor photometric axis \citep[Figure\,D.49 in][]{mky2020}. Due to the inclination, the dust structure looks like an eyebrow. 
In the MATLAS survey, NGC\,5485 is classified as having streams, disturbed outer isophotes, prominent dust lanes, and an unsure detection of shells \citep{matlas20}.

\subsection*{NGC\,5557} 

NGC\,5557, Fig.\,\ref{fig:lsma}, is another rich shell galaxy with around nine shells visible in the MATLAS image.
There are more tidal features of irregular shapes found up to about 13\arcmin{} (160\,kpc) from the center \citep[see also][]{duc11}.
In the MATLAS survey, NGC\,5557 is classified as having shells, tails, and disturbed outer isophotes \citep{matlas20}.

\subsection*{NGC\,6173} 

NGC\,6173 is a BCG in Abell\,2197 \citep{bcg95} with an outstanding Type\,I shell system. 
Around nine shells are visible in the Legacy Surveys image, Fig.\,\ref{fig:ls7}.

\subsection*{NGC\,7252 (Arp\,226)} 

NGC\,7252, Fig.\,\ref{fig:7252}, is a famous merger-remnant galaxy, also known under the nickname “Atoms for peace.”
The galaxy is equipped with two long tails, several loops, and many (more than ten) shells. 
The HST image also reveals a dusty disk in the inner parts. 

The VLT-VIMOS integral-field spectrograph observations uncover complex stellar kinematics with a clear prolate-rotating component, while the gas shows precise oblate rotation in the inner parts and prolate rotation in the outer parts \citep{wea18}.
A major merger of two disk galaxies is a widely accepted scenario for the origin of NGC\,7252 \citep{sch82,sch83,br82,br91,mbr93,hmc95}.
Combining the information from the on-going and past star formation and the gas and stellar morphology and kinematics with the results from theoretical studies, \cite{wea18} derived a probable time of the final coalescence of the merger to be within the last 200\,Myr.

\subsection*{PGC\,018579 (AM\,0609-331, ESO\,364-IG\,042)} 

As seen in the Legacy Surveys image, Fig.\,\ref{fig:ls7}, PGC\,018579 is a galaxy in an interaction -- it creates a contact pair with 2MASX\,J06110577-3318037. 
In the north, there is a hint of a faint shell around the companion.

\subsection*{PGC\,021757 (LSBC\,F560-04)} 

PGC\,021757 is a BCG with an exemplary Type\,I shell system.
In the Legacy Surveys image, Fig.\,\ref{fig:ls7}, at least nine shells are visible.

\section{Discussion} \label{sec:dis}

We detected, often multiple, signs of interaction in all 19 prolate rotators in our sample. 
Here, we compare the significance of the detection with the rate of interaction signs in the MATLAS sample. 

\subsection{Detectability in MATLAS}  \label{sec:sbl}

Only four of our prolate rotators are also in MATLAS. 
These four galaxies were also classified as tidally disturbed under the MATLAS classification \citep{matlas20}. 

Most of the galaxies in our sample have images in the Legacy Surveys.
We can safely say that those interaction signs detected in Legacy Surveys images would be detected if the galaxy was a part of the MATLAS sample, see Fig.\,\ref{fig:lsma}. 
The MATLAS images have a deeper surface-brightness limit, and the survey was optimized for detecting extended low-surface-brightness structures.
This is paramount in discovering the signs of past galaxy interactions. Without optimization, many signs of interaction can escape detection.

In some cases, especially for galaxies with larger angular sizes, the Legacy Surveys images can show the shape of the outer halo worse than the SDSS images (see NGC\,4874, Fig.\,\ref{fig:lsma}), even though the Legacy Surveys generally have a better surface-brightness limit than SDSS. 
This discrepancy is due to the artifacts at the borders of individual images/exposures in the Legacy Surveys and the local background subtraction that often suppresses the outer halo of the galaxies.

Five other prolate rotators in our sample do not have sufficiently good images in the Legacy Surveys.  
The three prolate rotators located in the Virgo cluster were examined using the Burrell Schmidt Deep Virgo Survey images.
As stated in \cite{mih17}, the surface-brightness limit of these images is 29.5 (28.5) mag\,arcsec$^{-2}$ in the $B$ ($V$) band, which is comparable with MATLAS. 
For NGC\,1052, we took advantage of the published image from the HERON survey with a similar surface-brightness limit of 28.5\,mag\,arcsec$^{-2}$ in the $r$ band \citep{mu19}.
NGC\,4589 is assessed in the image obtained with the Milankovi\'c telescope (for more details, see Appendix\,\ref{apx:obs}). 
From our experience, substructures in galaxies in the images from this telescope in $L$-band, taken with a comparable exposure time as the NGC\,4589 image, have similar visibility as in MATLAS \citep[Vudragovi{\'c} et al., in prep.; see also][]{mul19,bil20dg}.

NGC\,7252 is not included in the Legacy Surveys at all.
The image presented in the left panel of Fig.\,\ref{fig:7252} was taken by the Wide Field Imager on the MPG/ESO 2.2-metre telescope at ESO’s La Silla Observatory. 
The photometric depth of this image is comparable with Legacy Surveys images.

For two prolate rotators from our sample (NGC\,4589 and NGC\,4874), we also show the residual images (i.e. images after the model subtraction) to emphasize the detected substructures or reveal additional ones. However, all the galaxies have signs of tidal disturbance visible without the model subtraction.
In summary, all 19 prolate rotators in our sample have signs of galaxy interaction found in the images (without the model subtraction) that are shallower or have a similar depth compared with the MATLAS images.
We can confidently conclude that all 19 prolate rotators would be classified as tidally disturbed if they were a part of the MATLAS survey.

\subsection{Environment density}  \label{sec:env}

The deep imaging of the complete volume and magnitude-limited ATLAS$^{3{\rm D}}$ sample was divided into two groups: 
(1) the Next Generation Virgo Cluster Survey (NGVS) covering the full Virgo Cluster area \citep{ngvs1}, and 
(2) the MATLAS sample covering the rest of the ATLAS$^{3{\rm D}}$ galaxies \citep[][see also \citealt{duc13,duc20}]{duc15}. 
Thus, the MATLAS sample consists of galaxies only from environments with a relatively low galaxy density, while some of our prolate rotators reside in clusters.

Generally, the tidal structures are less abundant among the cluster galaxies \citep{tal09}. 
Especially shells, which occur in most of our prolate rotators, are known to occur significantly more in environments with low galaxy density \citep{mc83,red96,col01}. 
On the other hand, the preliminary results of NGVS show no significant difference in the frequency of tidal disturbance found in the NGVS and MATLAS galaxies (Duc, private communication). 
In summary, by neglecting the effect of the environment, we do not artificially increase the significance of our results, rather the opposite.

\subsection{Statistical significance of the results}  \label{sec:sig}

If the probability that a randomly chosen galaxy exhibits interaction signs is $p$, then the probability to find $k$ such galaxies in a sample of $N$ galaxies follows the binomial distribution:
\begin{equation}
P(k|p) = \binom{N}{k}p^k(1-p)^{N-k}.
\end{equation}
By Bayes' theorem and the law of total probability, the posterior  probability density function of $p$ given the observation of $k$ such galaxies in the sample is 
\begin{equation}
P(p|k) = \frac{P(k|p)P(p)}{P(k)} = \frac{P(k|p)P(p)}{\int_0^1P(k|p)P(p)\mathrm{d}p},
\end{equation}
where $P(k)$ is the prior probability of observing $k$ such galaxies in the sample, and $P(p)$ is the prior probability density function of $p$, which we assume  to be constant. 
The probability of finding $l$ such galaxies by chance in another sample of $M$ galaxies can be found by the application of the law of total probability:
\begin{equation}
\begin{aligned}
P(l) &= \int_{0}^{1} \binom{M}{l}p^l(1-p)^{M-l}\,P(p|k)\,\mathrm{d}p = \\
&=  \frac{N+1}{M+N+1}\frac{\binom{N}{k}\binom{M}{l}}{\binom{M+N}{k+l}}.
\end{aligned}
\end{equation}
and the probability to find at least $l$ such galaxies is $P(\geq l)=\sum_{l^\prime=l}^{M}P(l^\prime)$.

We take MATLAS as our reference sample.
The frequency of tidal features is known to increase with the host galaxy mass \citep[e.g.,][]{at13,matlas20}. 
The MATLAS sample consists of ETGs with $9.5<$\,log$(M_{\rm JAM})<11.8$. For galaxies with log$(M_{\rm JAM})>11$, the incidence of shells and streams increases about 1.7 times.
All but two galaxies in our sample of prolate rotators have the dynamical mass (or its approximate estimate; see Sect.\,\ref{sec:mass} and Column (7) in Table\,\ref{tab:prs}) log$(M_{\rm JAM})>11$. 
Thus, we are going to compare the frequency of interaction signs with the frequency among the 35 ETGs with log$(M_{\rm JAM})>11$ in MATLAS \citep[the entry “Any tidal disturbance” in Table\,7 in][]{matlas20}.
Out of the 35 ETGs, 22 are classified as tidally disturbed (including seven uncertain detections). 
That means we have $N=35$ and $k=22$ for the reference sample and $l=M=19$ for our sample of prolate rotators, leading to the chance probability of the observation equal to 0.00087. 
That proves the strong connection between the prolate rotation and the morphological signs of the galaxy interactions.

\subsection{Overabundance of shells} \label{sec:shells}

Here, we perform the same analysis as we did in Sect.\,\ref{sec:sig} for the signs of any tidal disturbance, but restricted only to shells.
The presence of multiple shells in the galaxy is most likely an indication of a significant merger undergone by the host galaxy.
We count only galaxies where multiple (at least three) shells are visible.
We did not include galaxies with an insufficient number of shells or an unclear detection of them (e.g., NGC\,647, 1052, 4261, 4486, 5485). 
We also did not count NGC\,4874 with multiple shells visible only in the residual image, and residual images were not part of the MATLAS classification procedure. 
Ten prolate rotators satisfy our criteria: NGC\,810, 1222, 2484, 2783, 4406, 5216, 5557, 6173, 7252, and PGC\,021757.

The reference sample is again the 35 MATLAS ETGs with log$(M_{\rm JAM})>11$.
Among them, there are eight ETGs classified as having or likely having shells, including all three of our prolate rotators that are in MATLAS in that mass bin. 
The fourth prolate rotator, NGC\,1222, is also classified as having shells, but it has log$(M_{\rm JAM})<11$.
From these eight ETGs, five (including one of our prolate rotators) satisfy our criteria: NGC\,680, 3613, 3613, 4382, 5557.
That means we have $N=35$ and $k=5$ for the reference sample; $M=19$ and $l=10$ for our sample of prolate rotators, leading to the chance probability of the observation equal to 0.0033. 
The significance of the connection between multiple shells and prolate rotation is thus lower than for tidal disturbances in general but still very high.

\subsection{Interpretation} \label{sec:itp}

The fact that a tidal feature is visible in a galaxy means that the related galaxy interaction (presumably merger) happened relatively recently, within the last few gigayears.
The overabundance of the interaction signs in prolate rotators has several possible explanations:
\setlist[enumerate,1]{label=(\arabic*),ref=\arabic*}
\begin{enumerate}
\item Prolate rotation is created in mergers, but generally, the rotation is not stable over a long time. Therefore, we only see prolate rotators formed in recent mergers.
\item Prolate rotation is created in mergers, and it is stable over a long time, but ancient mergers are not suitable for the production of prolate rotation. 
In such a case, we would, again, observe only prolate rotators formed in recent mergers.
Ancient mergers are typically considerably more gas-rich, even compared to recent wet mergers.
This condition can, in principle, lead to a quick renewal of oblate rotation after the merger.  
This outline is similar to the evolution of ETGs in Illustris reported by \cite{illfrsr}, where slow rotators are created during the last 8\,Gyr of the simulation by the combination of mergers and the inability to regain spin. 
This scenario requires the lifetime of tidal features to be, in most cases, comparable or longer than the epoch in which the mergers are sufficiently gas poor.
If the lifetime of tidal features were significantly shorter than the epoch, we would expect at least some of the prolate rotators to exhibit no signs of interactions.
\item Prolate rotation can be (in some or all cases) formed a long time ago, but preferentially in environments with frequent galaxy interactions.
In such a scenario, the hosts can contain signs of galaxy interaction that are not directly related to the origin of the prolate rotation. 
Prolate rotation could have been created either in mergers (the merger can be ancient in this scenario) or by another mechanism that has to occur preferentially in environments with frequent galaxy interactions. 
In this case, prolate rotation has to be able to survive through the event that created the tidal features that are currently observed in prolate rotators.
Support for this scenario comes from the fact that some of the prolate rotators are currently clearly experiencing galaxy interactions (PGC\,018579 and NGC\,5216) or are likely part of ongoing interactions while having signs of past, possibly unrelated, interactions (M86, M87, NGC\,1052, NGC\,4365, NGC\,4589, NGC\,4874, and NGC\,6173).
\end{enumerate}

We note that two or all three scenarios, or some of their aspects, can take place simultaneously in the Universe. 
The first two points are consistent with the large-scale cosmological hydrodynamical simulation Illustris, where prolate rotators are mostly created in major mergers within the last 6\,Gyr of the simulation \citep{illprol}. 
Such mergers are expected to produce tidal features that should be, in the majority of cases, visible in deep imaging surveys like MATLAS. 
However, even in this scenario, some of the interaction signs observed in prolate rotators may not be related to the origin of the prolate rotation.

\section{Conclusions}

We collected a sample of 19 nearby ETGs with convincing prolate rotation from ATLAS$^{3{\rm D}}$, CALIFA, and the literature. 
We inspected their deep optical images for signs of galaxy interactions\,--\,18 in the archival images and one in an image newly-obtained with Milankovi\'c telescope.
We found tidal tails, shells, or asymmetric/disturbed stellar halos in all 19 prolate rotators.
We compared this with the tidal disturbance frequency among the galaxies with a similar mass range drawn from the MATLAS sample -- a volume-limited general sample of nearby ETGs. 
We verified that the MATLAS survey has a similar or better surface-brightness limit than the images of the prolate rotators. 
We found that the chance probability of observing 19 out of 19 ETGs with interaction signs is only 0.00087. 

There is also a strong but less significant connection between the prolate rotation and the presence of multiple shells in the host galaxy, with the chance probability of the observation equal to 0.0033.
Our results agree with the Illustris large-scale cosmological hydrodynamical simulation, where prolate rotators are predominantly formed in major mergers during the last 6\,Gyr of the simulation. 

In addition, we report a discovery of a shell in the HST archival images of NGC\,7052. 
The galaxy is a known prolate rotator, but it could not be included in our sample because the available images do not qualify as comparable with MATLAS images.

\begin{acknowledgements}

We thank the referee for the quick and positive review. 
We are grateful to Christopher Mihos for providing the image of NGC\,4365 and to Michael Rich for providing the image of NGC\,1052.

We acknowledge the support from the Polish National Science Centre under the grant 2017/26/D/ST9/00449 (IE and MB).

A.V. acknowledges the financial support of the Ministry of Education, Science and Technological Development of the Republic of Serbia (MESTDRS) through the contract No. 451-03-9/2021-14/200002 and the financial support by the European Commission through project BELISSIMA (BELgrade  Initiative  for Space  Science,  Instrumentation  and  Modelling  in Astrophysics,  call  FP7-REGPOT-2010-5,  contract  No.  256772), which was used to procure the Milankovi\'c 1.4 meter telescope with the support from the MESTDRS.

The Legacy Surveys consist of three individual and complementary projects: the Dark Energy Camera Legacy Surveys (DECaLS; NOAO Proposal ID \# 2014B-0404; PIs: David Schlegel and Arjun Dey), the Beijing-Arizona Sky Survey (BASS; NOAO Proposal ID \# 2015A-0801; PIs: Zhou Xu and Xiaohui Fan), and the Mayall z-band Legacy Surveys (MzLS; NOAO Proposal ID \# 2016A-0453; PI: Arjun Dey). DECaLS, BASS and MzLS together include data obtained, respectively, at the Blanco telescope, Cerro Tololo Inter-American Observatory, National Optical Astronomy Observatory (NOAO); the Bok telescope, Steward Observatory, University of Arizona; and the Mayall telescope, Kitt Peak National Observatory, NOAO. The Legacy Surveys project is honored to be permitted to conduct astronomical research on Iolkam Du'ag (Kitt Peak), a mountain with particular significance to the Tohono O'odham Nation.
NOAO is operated by the Association of Universities for Research in Astronomy (AURA) under a cooperative agreement with the National Science Foundation.
This project used data obtained with the Dark Energy Camera (DECam), which was constructed by the Dark Energy Survey (DES) collaboration. Funding for the DES Projects has been provided by the U.S. Department of Energy, the U.S. National Science Foundation, the Ministry of Science and Education of Spain, the Science and Technology Facilities Council of the United Kingdom, the Higher Education Funding Council for England, the National Center for Supercomputing Applications at the University of Illinois at Urbana-Champaign, the Kavli Institute of Cosmological Physics at the University of Chicago, Center for Cosmology and Astro-Particle Physics at the Ohio State University, the Mitchell Institute for Fundamental Physics and Astronomy at Texas A\&M University, Financiadora de Estudos e Projetos, Fundacao Carlos Chagas Filho de Amparo, Financiadora de Estudos e Projetos, Fundacao Carlos Chagas Filho de Amparo a Pesquisa do Estado do Rio de Janeiro, Conselho Nacional de Desenvolvimento Cientifico e Tecnologico and the Ministerio da Ciencia, Tecnologia e Inovacao, the Deutsche Forschungsgemeinschaft and the Collaborating Institutions in the Dark Energy Survey. The Collaborating Institutions are Argonne National Laboratory, the University of California at Santa Cruz, the University of Cambridge, Centro de Investigaciones Energeticas, Medioambientales y Tecnologicas-Madrid, the University of Chicago, University College London, the DES-Brazil Consortium, the University of Edinburgh, the Eidgenossische Technische Hochschule (ETH) Zurich, Fermi National Accelerator Laboratory, the University of Illinois at Urbana-Champaign, the Institut de Ciencies de l'Espai (IEEC/CSIC), the Institut de Fisica d'Altes Energies, Lawrence Berkeley National Laboratory, the Ludwig-Maximilians Universitat Munchen and the associated Excellence Cluster Universe, the University of Michigan, the National Optical Astronomy Observatory, the University of Nottingham, the Ohio State University, the University of Pennsylvania, the University of Portsmouth, SLAC National Accelerator Laboratory, Stanford University, the University of Sussex, and Texas A\&M University.
BASS is a key project of the Telescope Access Program (TAP), which has been funded by the National Astronomical Observatories of China, the Chinese Academy of Sciences (the Strategic Priority Research Program "The Emergence of Cosmological Structures" Grant \# XDB09000000), and the Special Fund for Astronomy from the Ministry of Finance. The BASS is also supported by the External Cooperation Program of Chinese Academy of Sciences (Grant \# 114A11KYSB20160057), and Chinese National Natural Science Foundation (Grant \# 11433005).
The Legacy Surveys team makes use of data products from the Near-Earth Object Wide-field Infrared Survey Explorer (NEOWISE), which is a project of the Jet Propulsion Laboratory/California Institute of Technology. NEOWISE is funded by the National Aeronautics and Space Administration.
The Legacy Surveys imaging of the DESI footprint is supported by the Director, Office of Science, Office of High Energy Physics of the U.S. Department of Energy under Contract No. DE-AC02-05CH1123, by the National Energy Research Scientific Computing Center, a DOE Office of Science User Facility under the same contract; and by the U.S. National Science Foundation, Division of Astronomical Sciences under Contract No. AST-0950945 to NOAO.

Based on observations made with the NASA/ESA Hubble Space Telescope, obtained from the data archive at the Space Telescope Science Institute. STScI is operated by the Association of Universities for Research in Astronomy, Inc. under NASA contract NAS 5-26555.

Funding for the Sloan Digital Sky Survey IV has been provided by the Alfred P. Sloan Foundation, the U.S. Department of Energy Office of Science, and the Participating Institutions. SDSS acknowledges support and resources from the Center for High-Performance Computing at the University of Utah. The SDSS web site is www.sdss.org.
SDSS is managed by the Astrophysical Research Consortium for the Participating Institutions of the SDSS Collaboration including the Brazilian Participation Group, the Carnegie Institution for Science, Carnegie Mellon University, the Chilean Participation Group, the French Participation Group, Harvard-Smithsonian Center for Astrophysics, Instituto de Astrofísica de Canarias, The Johns Hopkins University, Kavli Institute for the Physics and Mathematics of the Universe (IPMU) / University of Tokyo, the Korean Participation Group, Lawrence Berkeley National Laboratory, Leibniz Institut für Astrophysik Potsdam (AIP), Max-Planck-Institut für Astronomie (MPIA Heidelberg), Max-Planck-Institut für Astrophysik (MPA Garching), Max-Planck-Institut für Extraterrestrische Physik (MPE), National Astronomical Observatories of China, New Mexico State University, New York University, University of Notre Dame, Observatório Nacional / MCTI, The Ohio State University, Pennsylvania State University, Shanghai Astronomical Observatory, United Kingdom Participation Group, Universidad Nacional Autónoma de México, University of Arizona, University of Colorado Boulder, University of Oxford, University of Portsmouth, University of Utah, University of Virginia, University of Washington, University of Wisconsin, Vanderbilt University, and Yale University.

This research has made use of the NASA/IPAC Extragalactic Database, which is funded by the National Aeronautics and Space Administration and operated by the California Institute of Technology.

This research has made use of the SIMBAD database, operated at CDS, Strasbourg, France

\end{acknowledgements}


\bibliographystyle{aa}
\bibliography{prolrot} 


\appendix

\section{Shell in NGC\,7052}  \label{sec:n7052}

\begin{figure*} 
\centering
\includegraphics[width=\hsize]{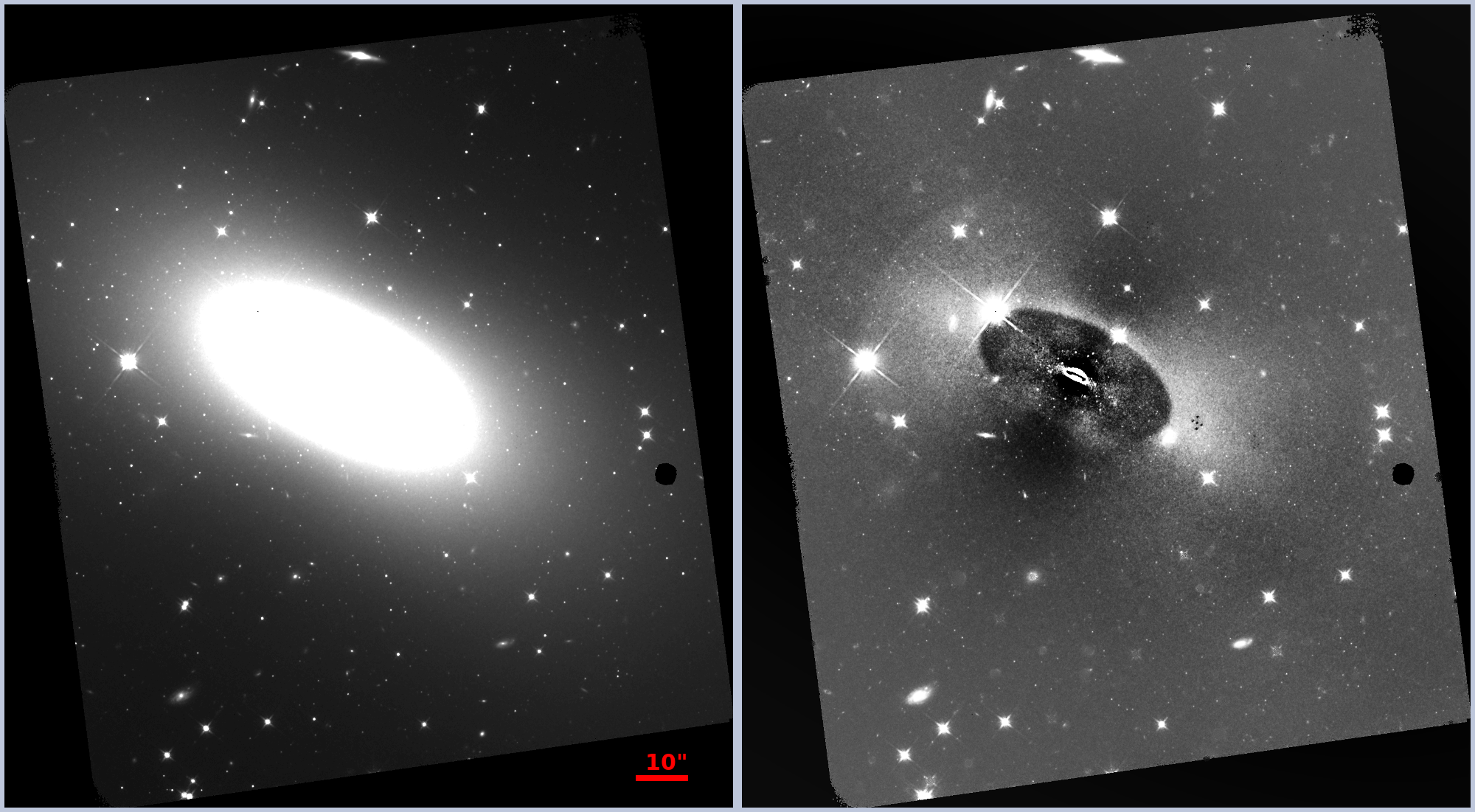}
\caption{
Images of NGC\,7052, a prolate rotator that is not part of our sample. Left: the original HST image; right: the residual image of the one on the left. The image in the right panel has been obtained by combining two different models, \textsc{iraf}'s 1D ellipse for inner parts and \textsc{galfit} for outer parts, see Appendix\,\ref{sec:n7052} for more details. Both panels have the same field of view. North is up, east is to the left.
The newly discovered shell, especially clearly visible in the residual image, is situated about 42\arcsec{} in the northeast direction.
\label{fig:7052}
}
\end{figure*}

NGC\,7052, Fig.\,\ref{fig:7052}, is an isolated field galaxy with no nearby companion. 
It is a known prolate rotator \citep{wa88} that is not a part of our sample. 
The galaxy was observed with HST, but generally, HST probes galaxies at different scales than CFHT MegaCam. 
Thus the HST images do not qualify as comparable with MATLAS and we cannot include NGC\,7052 in the sample, but we inspected the images anyway. 

We used the HST/WFC3 F110W archival data obtained on Jun\,14, 2016 (exposure time 2496\,s; PI: John Blakeslee). 
We found a previously unreported shell in the northeast direction, about 42\arcsec{} from the center along the major photometric axis. 
The galaxy also shows strong outer boxy isophotes \citep{lau85,bdm88,gs93,dej94,vk99} indicating a possible presence of more unresolved outer shells.
Additionally, there is a large central dust disk with a diameter of 1\,kpc (4.0\arcsec) oriented along the optical major axis \citep{nie90,dej96}.

To improve the visibility of the shell, we modeled the galaxy using the same method as for NGC\,4589. However, \textsc{iraf}'s 1D ellipse fitting algorithm cannot overcome the boxy outer regions and results in an X-shaped residual. Therefore, we used the \textsc{galfit} program \citep{galfit2010} with two S\'ersic and one exponential component. As a result, the \textsc{galfit} finds the boxy outer shape's presence; the parameter C0 is approximately 0.24. The \textsc{galfit} cannot model the center since there is an intense dust lane. Therefore, we used a mixed model: Having matched the scales of the models, we replaced the inner regions of the \textsc{galfit} model with that of the \textsc{iraf} model. To remove the sharp transition, we smoothed the mixed model with a Gaussian function. The right panel of Fig.\,\ref{fig:7052} shows the residual image obtained by using the mixed model.

Such a shell should be detectable in the MATLAS survey. 
The shell in NGC\,7052 is located 42\arcsec{} from the center and it is visible even without any special image processing. 
In \cite{bil16}, a famous shell galaxy NGC\,3923 was analyzed using HST and MegaCam images.
MegaCam is the same instrument that was used for the MATLAS survey and the images of  NGC\,3923 have a similar depth and observation strategy as MATLAS. 
Even the innermost (11\arcsec{} from the center) shells of NGC\,3923 and all the shells that are visible without a special image processing, are detected in both HST and MegaCam images.
However, the HST images probe a different range of galactocentric distances, thus the HST images cannot be used as a source for the systematic search for tidal disturbance when compared with the MATLAS results.

\section{NGC\,4589 observations} \label{apx:obs}
	
NGC\,4589 was observed on 15/09/2020 with the 1.4-meter Milankovi\'c telescope equipped with an Andor IKONL CCD camera mounted at the Astronomical Station Vidojevica (Serbia). Total of 205 frames, each exposed for 100\,s in the $L$-band, was taken with the dithering of 500 pixels (195\arcsec), summing up to the 5.7h integrated on-source exposure time. The field-of-view with the focal reducer of 13\arcmin.3 $\times$ 13\arcmin.3 was enlarged by random dithering to the 23\arcmin.8 $\times$ 23\arcmin.9 in the final co-add image. 	
		
Raw images were reduced using the pipeline for the Milankovi\'c telescope \citep{mul19} optimized for the detection of low-surface-brightness features. This pipeline creates the sky flat using largely dithered images. However, dithering of 500 pixels was not large enough to ensure flatness of the background in the final co-add image. So, the sky flat was created median combining images from which the galaxy was removed. Prior to the reduction, galaxy was modelled and subtracted from all raw images with the Galfit code \citep{2010AJ....139.2097P}, where, in addition to other objects, the central part of the galaxy was masked to enable Galfit to better model the outer parts of the galaxy surface brightness profile. This intervention substantially improved the final co-add in terms of the flatness of the background. The final co-add image was created using SWarp software  \citep{2010ascl.soft10068B} median combining all calibrated frames for which WCS solution was obtained using Astrometry software \citep{2010AJ....139.1782L}.

  \end{document}